\documentclass[pra,twocolumn,showpacs,preprintnumbers,amsmath,amssymb,superscriptaddress]{revtex4}
\usepackage{revsymb}
\usepackage[cp1250]{inputenc}   
\usepackage[T1]{fontenc}
\usepackage{amssymb,amsmath,amscd,amsfonts}
\usepackage[dvips]{graphicx}
\usepackage{dcolumn}
\usepackage[FIGTOPCAP,raggedright,nooneline,bf,footnotesize]{subfigure}
\usepackage[hypertex]{hyperref}
\usepackage{indentfirst}
\usepackage{multirow}
\usepackage[usenames]{color}
\newcommand{\mathsym}[1]{{}}

\newcommand{\V}[1]{\mathbf{#1}}
\newcommand{\kp}[1]{\V{k}_{#1 \perp}}

\renewcommand{\vec}[1]{\mathbf{#1}}
\newcommand{\mc}[1]{\mathcal{#1}}
\newcommand{\allarg}{\kp{s},\omega_s;\kp{i},\omega_i}
\newcommand{\atemp}{A_{\text{sp}}}
\newcommand{\asp}{ u_p}
\newcommand{\dd}{\boldsymbol{\Delta_0}}
\newcommand{\zero}[1]{\mathbb{O}_#1}

\renewcommand{\eqref}[1]{Eq.~(\ref{#1})}
\newcommand{\figref}[1]{Fig.~\ref{#1}}

\newcommand{\appref}[1]{Appendix~\ref{#1}}

\newcommand{\de}{\text{d}}
\newcommand{\rozw}{\mathbf{D}}
\newcommand{\blroz}{\mathbf{d}}

\newcommand{\spectfil}{\Lambda}
\newcommand{\nf}{\mc{N}}
\newcommand{\sr}{\Psi_{i}}
\newcommand{\dm}{\rho_i(\omega_i,\omega_i')}

\begin{document}

\title{Density matrix of a single photon produced in parametric down conversion derived}
\author{Piotr Kolenderski}
\homepage{http://www.fizyka.umk.pl/~kolenderski}
\affiliation{Institute of
Physics, Nicolaus Copernicus University, Grudziadzka 5, 87-100
Toru{\'n}, Poland}
\author{Wojciech Wasilewski}
\affiliation{Institute of Experimental Physics, Warsaw University,
Ho{\.z}a 69, 00-681 Warsaw, Poland}
\affiliation{Niels Bohr Institute, University of Copenhagen, DK 2100, Denmark}
\affiliation{QUANTOP, Danish National Research Foundation Center for
Quantum Optics}
\date{\today}

\date{\today}
\begin{abstract}
    We introduce an effective numerical method of density matrix determination of fiber coupled single photon generated in process of spontaneous parametric down conversion in type I non-collinear configuration. The presented theory has been successfully applied in case of source exploited  in Phys. Rev. Lett. \textbf{99}, 123601 (2007) to demonstrate the experimental characterization of spectral state of single photon.
\end{abstract}

\maketitle

\section{Introduction}
The source of single photons is a prerequisite for implementation of
linear optical quantum information processing schemes
\cite{Knill2001,OBrien2003,Kok2007,Lu2007}, quantum teleportation
\cite{Marcikic2003} and quantum cryptography protocols
\cite{Gisin2002}. Altough deterministic single photon sources has been
recently developed
\cite{Kurtsiefer2000,Beveratos2002,Santori2004,Rabeau2007}, most of
the fundamental experiments still utilize the process of Spontaneous
Parametric Down Conversion (SPDC).  In this phenomena pairs of
daughter photons are produced in a pumped nonlinear crystal and
typically coupled into single mode fibers (SMFs). SPDC provides unique
ease of shaping spectral and spatial properties of generated
nonclassical light. Detection of one photon, conventionally called
signal, heralds the presence of the other, called idler
\cite{rarity1997}. Its statistical properties are satisfactory for
post-selection type experiments as long as the pump power is low
enough to make multiple pair generation events negligible
\cite{Wasilewski2008}. However, the idler photon is typically prepared
in a mixed state \cite{URen2005} with respect to the temporal degree
of freedom. Only by careful spectral filtering or precise adjustment
of collecting setup one can produce pure wave packets
\cite{Riedmatten2003, Kaltenbaek2006, Mosley2008, Kolenderski2009}.
Furthermore, coupling into SMFs decreases significantly the total
number of useful heralded photons. In a typical experimental scenario
coupling only one photon out of the pair is significantly more
probable than coupling both photons. This comes at an expense of
loosing sub-poissonian statistics of the counts but is beneficial for
a certain test experiments \cite{Wasilewski2007} or characterising
vital characteristics of the source.

We report here an effective numerical method of determination of the
state of single photon generated in process of type I SPDC in
configuration in which idler photon is coupled into SMF and the signal
is disregarded, adopted from the results of \cite{Kolenderski2009}.
This scheme has been previously used to demonstrate the method of
characterisation of a single photon state in Ref.
\cite{Wasilewski2007}. The theoretical method is based on an
observation that optical fiber define a relatively narrow range of
directions that need to be included in the calculations. This
justifies the paraxial approximation, which makes a substantial
portion of the problem tractable analytically. In consequence we are
able to derive the expression for the spectral density matrix of idler
photon $\dm$ propagating in SMF.Finally we present a numerical calculations of $\dm$ for a typical experimental settings. Both obtained results --- highly
simplified analytical assessment of the single photon characteristics
and an efficient method of calculating its full density matrix $\dm$ ---
can provide a valuable insight in a future experiments with single
photons obtained from SPDC.

\section{Theoretical model}

Below we will develop a set of effective approximations for calculating the reduced density matrix $\dm$ in a single arm of typical fiber coupled single-pass parametric down-conversion source. We start by writing down the full biphoton wavefunction in the perturbative approximation. We describe the coupling of one of the photons into a SMF and we trace over the other photon. This way we obtain an involved, multidimensional integral for the reduced density matrix $\dm$. At this stage we point out, that by judicious reordering of the integrals and performing a paraxial approximation this expression can be largely simplified and cast into a form which is easy to evaluate.

\begin{figure}
  \includegraphics[width=1\columnwidth]{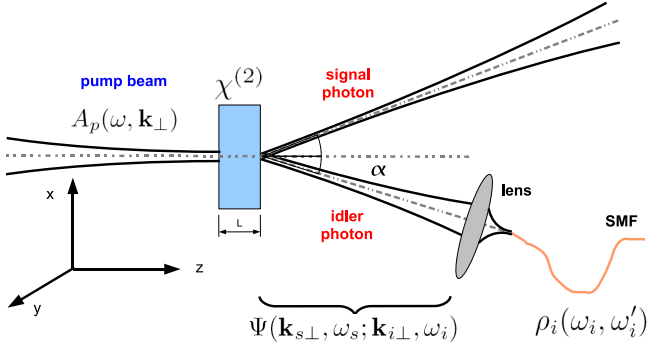}\\
  \caption{The setup is comprised of $\chi^{(2)}$ nonlinear crystal of length $L$, pumped by a gaussian beam of $w_p$ comprised of pules described by a spectral envelope function $A_{ps}$. The idler photon is collected to a single mode fiber SMF, provided it is emitted into a spatial mode of characteristic width $w_f$at the crystal.}
  \label{fig:setup}
\end{figure}

Let us begin by describing the typical source depicted in \figref{fig:setup}.
It  comprises the  nonlinear crystal of length $L$, pumped by a beam of ultrashort pulses centered around frequency $2\omega_0$. The elementary process taking place in nonlinear crystal is a decay of a pump photon into signal and idler photons. We chose to parameterize the interacting waves by their frequencies $\omega_p$, $\omega_s$, $\omega_i$ and transverse components of their wave vectors $\kp{p}$, $\kp{s}$, $\kp{i}$. The energy conservation principle is expressed in this parametrization by an equation $\omega_p=\omega_s+\omega_i$, while the momentum conservation requires that the transverse components of wave vectors match $\kp{p}=\kp{s}+\kp{i}$. Note that for each of the waves once $\kp{}$ and $\omega$ are given $k_z$ is fixed. Therefore an ideal match of the longitudinal wave vectors is in general impossible. We choose a perfect matching for pump photon of frequency $\omega_p=2\omega_0$ propagating along the $z$ axis ($\kp{p}=0$) and signal and idler photons of equal frequencies $\omega_s=\omega_i=\omega_0$ propagating symmetrically at the angle $\alpha$ with respect to $z$ axis. The propagation directions of both generated photons correspond to a transverse wave vectors $|\kp{s0}|=|\kp{i0}|=\omega_0/c\sin\alpha$ pointing in opposite directions. Thus a phase matching, which we consider here, corresponds to fulfilling the following criterion:
\begin{multline}\label{eq:PhaseMismatch}
    \Delta k_z(\kp{s0},\omega_{0};\kp{i0},\omega_{0})= \\ k_{pz}(\kp{s0}+\kp{i0},2\omega_0)-k_{sz}(\kp{s0},\omega_0)-k_{iz}(\kp{i0},\omega_0) =0
\end{multline}


Let us write down an expresion for the probability amplitude of generation two photons of frequencies $\omega_s$, $\omega_i$ and transverse wave vectors $\kp{s}$, $\kp{i}$ \cite{Rubin1997, Kolenderski2009}:
\begin{multline}\label{eq:FreeSpaceBWF}
    \Psi(\allarg)=\\
    \nf \int_{-L/2}^{L/2}\text{dz} A_p(\kp{s}+\kp{i},\omega_s+\omega_i)  e^{ i\Delta k_z(\allarg) z },
\end{multline}
where $\nf$ is normalization constant. The above formula expresses a fact that the probability amplitude of generating pair of photons characterized by $\kp{s},\omega_s$ and $\kp{i},\omega_i$ is proportional to the pump amplitude $A_p(\kp{s}+\kp{i},\omega_s+\omega_i)$. In turn the integration sums the contributions form the slices of the crystal perpendicular to $z$ axis, with exponential term representing propagation phases. Indeed, $\Delta k_z(\allarg) z$ is the phase acquired by the pump while propagating to the slice of interaction and by the photon pair while propagating towards the end of the crystal.

Now we must take a certain spatial shape of the pump and the fiber spatial mode, preferably leading to analytically integrable expressions. We assume there are no spatiotemporal correlation in the pump pulse and its amplitude is a product of spectral $\atemp(\omega)$ and spatial $\asp(\kp{})$ parts:
\begin{equation}\label{eq:PumpAmplidue}
    A_p(\omega,\kp{}) = \atemp(\omega) \asp(\kp{})
\end{equation}
We impose a spatial profile to be gaussian, corresponding to a beam of waist $w_p$: $$\asp(\kp{})=\frac{w_p}{\sqrt{\pi}}\exp\left( -\frac{w_p^2}{2}\kp{}^2 \right).$$
However, we still provide freedom of choice of the pump spectral amplitude $\atemp(\omega)$.

Next we find the wave function with respect to the fiber for an idler photon and with respect to the free space for the a signal photon. This can be done by projecting $\psi(\allarg)$ on the fiber mode profile. We approximate the spatial mode coupled in SMF $u_i(\kp{s})$ by a gaussian function centered around the phase matching direction $\kp{i0}$ corresponding to a beam of width $w_f$ \footnote{In the experiment this is the width of the beam on the crystal  one obtains by coupling the laser to the far end of the collecting fiber}:
\begin{equation}\label{eq:SMFModeProfile}
    u_i(\kp{i},\omega_i)=\frac{w_f}{\sqrt{\pi}} \exp\left(-\frac{w_f^2}{2}\left(\kp{i}-\kp{i 0}\right)^2 \right).
\end{equation}
Without loss of generality we assume the collecting optics and the fibers are in the $xz$ plane. Therefore the transverse wave vector of the idler photon is a function of its frequency $\omega_i$ in the direction of observation $\alpha$ is given by  $\kp{i0}=-\omega_i \sin(\alpha) \hat{x}/c$. Here $\hat x$ is a unit vector in the direction of $x$ axis. Thus the biphoton wave function of idler photon inside SMF and signal propagating in free space $\sr(\kp{s},\omega_s,\omega_i)$ is a partial overlap of the free space wave function $\Psi(\allarg)$ and the fiber mode profile $u_i(\kp{s})$:
\begin{multline}\label{eq:IdlerPhotonCoupledBWF}
\sr(\kp{s},\omega_s,\omega_i) =
    \nf \atemp (\omega_s+\omega_i)\int_{-L/2}^{L/2}\text{dz} \int \text{d}^2\kp{i} \\ u_i^*(\kp{i},\omega_i) \asp(\kp{s}+\kp{i}) e^{ i\Delta k_z(\allarg) z }.
\end{multline}

In the next step let us write an expression for the density matrix of the idler photon in the fiber $\dm$. It is obtained by taking a trace of the photon pair density matrix $\Psi^*(\allarg)\Psi(\kp{s}',\omega_s';\kp{i}',\omega_i')$ over the degrees of freedom of the signal photon $\omega_s$ and $\kp{s}$:
\begin{equation}\label{eq:DensityMatrixDef}
    \dm= \int\ \de\omega_s \de \kp{s}\ \Psi^*_{i}(\kp{s},\omega_s,\omega_i) \sr(\kp{s},\omega_s,\omega_i').
\end{equation}
Thus a general structure of the spectral density matrix $\dm$ after substitution of \eqref{eq:IdlerPhotonCoupledBWF} into \eqref{eq:DensityMatrixDef} is given by:
\begin{multline}\label{eq:DensityMatrixSymb}
    \dm\propto \int\  \de z \de z' \de\omega_s \atemp(\omega_s+\omega_i) \atemp(\omega_s+\omega_i') \times \\ \times \int \de \kp{s} \de \kp{i} \de \kp{i}'  \asp^* \asp'  u^*_i u_i' \exp(i \Delta k_z' z' -i \Delta k_z z),
\end{multline}
where the functions $\asp$, $u_i$ and $\Delta k_z$ are taken for $\kp{s}$, $\kp{i}$, $\omega_s$, $\omega_i$ and $\asp'$, $u_i'$ and $\Delta k_z'$ refer to the respective quantities taken at $\kp{s}$, $\kp{i}'$, $\omega_s$, $\omega_i'$.
A classical approach to evaluating this integral would be to first integrate analytically over the length of the crystal $z$ and $z'$ and then evaluate remaining 7 integrals numerically. However this would be extremely tedious computationally. Therefore we adopted another approach in order to alleviate the numerical effort.  We first perform the paraxial approximation of the phase mismatch exponent $i \Delta k_z' z' -i \Delta k_z z$. Then the integrals in second line of \eqref{eq:DensityMatrixSymb} can be evaluated analytically. Finally the remaining integrals over $z$, $z'$ and $\omega_s$ are computed numerically.

The paraxial approximation of the phase mismatch exponent $i \Delta k_z' z' -i \Delta k_z z$ is well justified as long as waists of the beam involved are much bigger than a wavelength. Mathematically this is equivalent to observing
that in the integral \eqref{eq:DensityMatrixSymb} the range of $\kp{i}$ is in fact restricted by the aperture of the collection optics described by  $u_i(\kp{s})$ to $\kp{i}\simeq\kp{i 0}\pm 1/w_f$, as seen in \eqref{eq:SMFModeProfile}. Then the range of perpendicular component of the signal wave vector $\kp{s}$ is also restricted because of finite wave vector spread of the pump. Direct calculation reveals  the range of approximately \footnote{See Sec.~IV in Ref.~\cite{Kolenderski2009}}: $\kp{s} \simeq \kp{s 0}\pm \sqrt{1/w_f^2+1/w_p^2}$, where $\kp{s0}=\omega_s \sin(\alpha) \hat{x}/c$. Those two facts enable us to approximate the phase mismatch $\Delta k_z(\allarg)$ around the center of both ranges
up to second order in $\kp{s}$ and $\kp{i}$:
\begin{multline}\label{eq:PhaseMismachApprox}
    \Delta k_z(\kp{s},\omega_s;\kp{i},\omega_i) :=
    \\
    k_{pz}(\kp{s}+\kp{i},\omega_s+\omega_i)-k_{sz}(\kp{s},\omega_s)-k_{iz}(\kp{i},\omega_i)\approx \\
    \Delta k_z(\kp{s0},\omega_{s};\kp{i0},\omega_{i}) +\rozw_{1}^T(\omega_s,\omega_i) \cdot\boldsymbol{\kappa} +{\boldsymbol{\kappa}}^T \rozw_{2}(\omega_s,\omega_i)\boldsymbol{\kappa},
\end{multline}
where for sake of brevity we introduced $\boldsymbol{\kappa}=(\kp{s}-\kp{s0},\kp{i}-\kp{i0})^T$ to denote the four component vector of deviations form the directions of phase matching. The expansion coefficients $\Delta k_z(\kp{s},\omega_s;\kp{i},\omega_i)$, $\rozw_{1}(\omega_s,\omega_i)$ and $\rozw_{2}(\omega_s,\omega_i)$ are given in the \appref{appendix:paraxial}. Note that for each pair of frequencies $\omega_s$ and $\omega_i$ we have in general different expansion coefficients.  This means we keep dispersion relations exact. This is motivated by the fact, that the spectrum of the single photons can easily span 100nm or more, which makes expansion of the $k$ vector as a function of frequency inaccurate.

With Taylor expansion given by \eqref{eq:PhaseMismachApprox} at hand, we can perform inner integrals in the expression for density matrix \eqref{eq:DensityMatrixSymb}. Let us rewrite the integrant of \eqref{eq:DensityMatrixSymb} using a vector
of deviations $\boldsymbol{\tilde \kappa}=(\kp{s}-\kp{s0},\kp{i}-\kp{i0},\kp{i}'-\kp{i0}')$ to shorten the notation:
\begin{multline}
    \asp^* \asp' u^*_i u_i' \exp(i \Delta k_z' z' -i \Delta k_z z)= \\ \exp\left(-\boldsymbol{\tilde \kappa}\vec{M}_2\boldsymbol{\tilde \kappa} +\vec{M}_1\boldsymbol{\tilde \kappa} +\vec{M}_0\right)
\end{multline}
where $\vec{M}_0$, $\vec{M}_1$ and $\vec{M}_2$ depend on frequencies $\omega_s$, $\omega_i$, $\omega'_i$ and $z$, $z'$. Their explicit form is given in \appref{appendix:M}. Now the integral in \eqref{eq:DensityMatrixSymb} can be partially evaluated and the density matrix $\dm$ is found to be:
\begin{multline}\label{eq:DensityMatrixFinal}
    \dm= |\nf|^2 \int \de\omega_s \atemp^*(\omega_s+\omega_i)\atemp(\omega_s+\omega_i')\times \\
    \int\limits_{-L/2}^{L/2} \de z\int\limits_{-L/2}^{L/2} \de z'\frac{\exp\vec{M}_0}{\sqrt{\det\vec{M}_2}}\exp\left(-\frac{1}{4}\vec{M}_1^T \vec{M}_2^{-1}\vec{M}_1\right).
\end{multline}

\section{Numerical results}
In order to proceed we fix the spectral shape of pump pulse $\atemp(\omega)$. We assume it be gaussian of duration $\tau_p$:
 \begin{equation}\label{eq:PumpSpectrum}
    \atemp(\omega)=\frac{\sqrt \tau_p}{\sqrt[4]{ \pi}} \exp\left(-\frac{\tau_p^2 }{2}(\omega-2\omega_0)^2\right)
\end{equation}
In typical experimental scenario photons are spectrally filtered in order to ameliorate its characteristics or reduce a noise. We can model spectral filtering effects multiplying  the density matrix by the the amplitude transition functions ${\spectfil(\omega')}$:
\begin{equation}
\rho_i(\omega,\omega') \rightarrow {\spectfil(\omega)}{\spectfil(\omega')}\rho_i(\omega,\omega')
\end{equation}
We approximate amplitude transmission of spectral filter by a Gaussian function defined in the following way:
\begin{equation}\label{eq:IF}
    \spectfil(\omega)=\exp\left(-2 \ln 2\frac{(\omega-\omega_0)^2}{\sigma^2}\right)
\end{equation}

\begin{figure}[ht]
    \begin{tabular}{cc}
      \subfigure[] {\includegraphics[width=0.46\columnwidth]{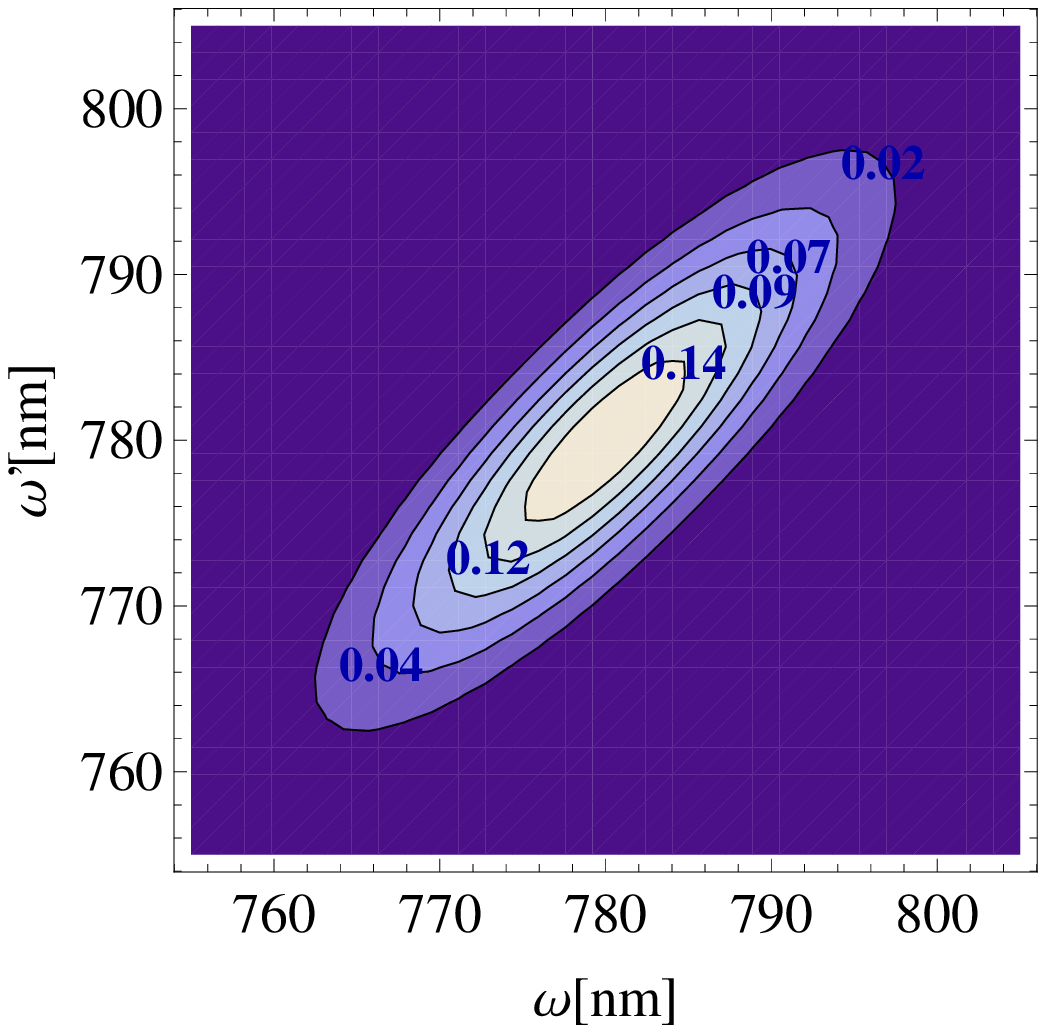}}&
      \subfigure[]            {\includegraphics[width=0.46\columnwidth]{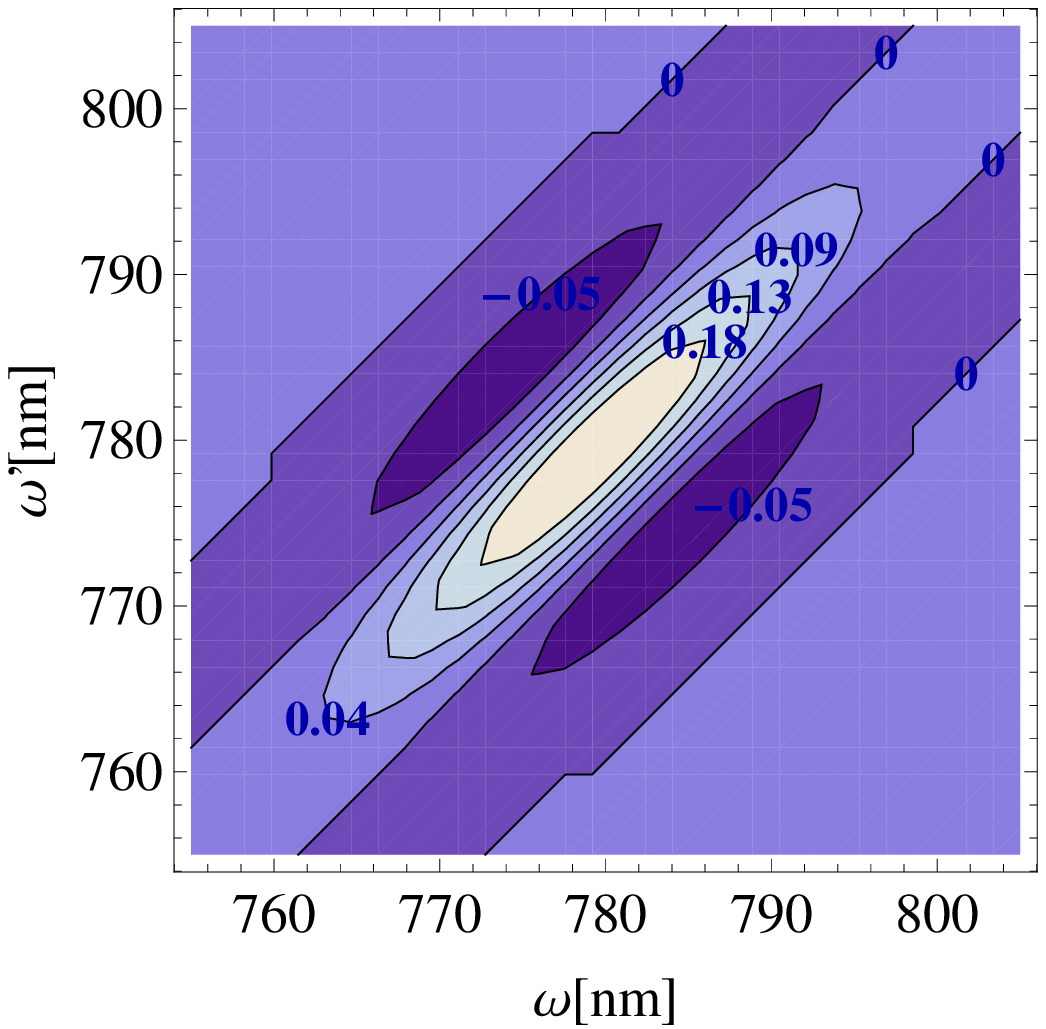}}\\
    \end{tabular}
  \caption{The spectral density matrix $\rho(\omega,\omega')$ of photons transmitted through 20nm spectral filter. The plots were generated for $L=1$mm long BBO crystal cut for $\theta_c=30^\circ$, pumped using gaussian beam of $w_p=100\mu$m waist. The pulse duration was set to $\tau_p=100$fs. The collecting optics and the single mode fiber was set to couple the gaussian mode of $w_f=100\mu$m. The observation angle was set to a) $\alpha=2.2^\circ$ (direction of perfect phase matching) and b) $\alpha=3.2^\circ$.}
  \label{fig:dm}
\end{figure}

Here we have calculated the spectral density matrix of photon generated within the typical experimental scenario, where the nonlinear crystal is pumped by a gaussian pulses \eqref{eq:PumpSpectrum}. Note that for real gaussian pulses density matrix is real.
\figref{fig:dm} shows the spectral density matrices $\rho(\omega,\omega')$ of photons generated in $L=1$mm long BBO crystal cut for $\theta_c=30^\circ$, pumped using gaussian pulsed beam of $w_p=100\mu$m waist and duration $\tau_p=100$fs. The collecting optics and the single mode fiber were set to couple the gaussian mode of waist  equal $w_f=200\mu$m. The spectral filters of $\sigma=20$nm were applied. Panel a) shows density matrix when the observation direction coincides with the phase matching direction $\alpha=2.2^\circ$, see \figref{fig:setup}. In panel b) the observation direction is set to $\alpha=3.2^\circ$.


\section{Conclusions}

We have derived a simple method of predicting the spectral density matrix $\dm$ of a photon generated in one arm of pulsed SPDC. The model allows to  reduce dramatically the computational effort while retaining the accuracy of results. Those results, by the virtue of their simplicity, may become very useful when engineering photon sources to meed a certain needs for the spectral characteristics.


The presented model has been successfully used to predict the measurement outcome in Ref.~\cite{Wasilewski2007, Kacprowicz2008}. Note that the model \eqref{eq:DensityMatrixFinal} may be used to predict the spectral density matrix in case of any type of spectral pumping.


\section{Acknowledgements}
This work has been supported by the Polish budget funds for scientific research projects in years 2005-2008 (Grant No. 1 P03B 011 29) and the European Commission under the Integrated Project Qubit Applications (QAP) funded by the IST directorate as Contract Number 015848.

\appendix
\section{Paraxial approximation} \label{appendix:paraxial}
The standard paraxial approximation is given by:
\begin{multline}\label{eq:PhaseMismachApprox}
    \Delta k_z(\kp{s},\omega_s;\kp{i},\omega_i) \approx
    \Delta k_z(\kp{s0},\omega_{s};\kp{i0},\omega_{i})+\\
    +\rozw_{1}^T(\omega_s,\omega_i) \cdot\boldsymbol{\kappa} +{\boldsymbol{\kappa}}^T \rozw_{2}(\omega_s,\omega_i)\boldsymbol{\kappa},
\end{multline}
where for sake of brevity we introduced $\boldsymbol{\kappa}=(\kp{s}-\kp{s0},\kp{i}-\kp{i0})^T$ to denote the four component vector of deviations form the directions of phase matching.
$\Delta k_z(\kp{s0},\omega_s;\kp{i0},\omega_i)$ is spatially constant term while:
\begin{equation}
    \rozw_{1}(\omega_s,\omega_i) = \left(\begin{array}{c} \blroz_s(\omega_s,\omega_i)\\ \blroz_i(\omega_s,\omega_i)
    \end{array}
\right)
\end{equation}
and
\begin{equation}
    \rozw_{2}(\omega_s,\omega_i) =\frac{1}{2}
    \left( \begin{array}{cc} \blroz_{ss}(\omega_s,\omega_i), & \blroz_{si}(\omega_s,\omega_i) \\ \blroz_{si}(\omega_s,\omega_i), & \blroz_{ii}(\omega_s,\omega_i) \end{array}\right),
\end{equation}
where we denote the blocks as:
\begin{equation}
    \blroz_{\mu}(\omega_s,\omega_i) = \left(\frac{\partial \Delta k_z}{\partial k_{\mu x}},\frac{\partial \Delta k_z}{\partial k_{\mu y}}\right),
    \end{equation}
and
\begin{equation}
    \blroz_{\mu\nu}(\omega_s,\omega_i) = \left(
    \begin{array}{cc} \displaystyle
    \frac{\partial^2 \Delta k_z}{\partial k_{\mu x} \partial k_{\nu x}}, & \displaystyle \frac{\partial^2 \Delta k_z}{\partial k_{\mu x} \partial k_{\nu y}} \\    \displaystyle \frac{\partial^2 \Delta k_z}{\partial k_{\mu y} \partial k_{\nu x}}, & \displaystyle
    \frac{\partial^2 \Delta k_z}{\partial k_{\mu y} \partial k_{\nu y}} \end{array}\right),
\end{equation}
In the above indexes $\mu, \nu = s,i$ refer to signal and idler photons.

\section{Definitions of density matrix coefficients}
\label{appendix:M}
The formula for spectral density matrix can be written down in the very compact form:
\begin{multline}
    \dm= |\nf|^2 \int \de\omega_s \atemp^*(\omega_s+\omega_i)\atemp(\omega_s+\omega_i')\times \\
    \int\limits_{-L/2}^{L/2} \de z\int\limits_{-L/2}^{L/2} \de z'\frac{\exp\vec{M}_0}{\sqrt{\det\vec{M}_2}}\exp\left(-\frac{1}{4}\vec{M}_1^T \vec{M}_2^{-1}\vec{M}_1\right)
\end{multline}
thanks to the following definitions:
\begin{eqnarray}
\displaystyle
\nonumber    \vec{M}_2&=&
        \frac{w_p^2}{2}
          \left(
            \begin{array}{ccc}
                2\openone_2 & \openone_2        & \openone_2 \\
                \openone_2  &(1+\frac{w_f^2}{w_p^2})\openone_2 & \zero{2}  \\
                \openone_2 & \zero{2}&(1+\frac{w_f^2}{w_p^2})\openone_2
            \end{array}
          \right)+
\\
&&
          -\frac{1}{2}\left(
            \begin{array}{ccc}
                i z' \blroz_{ss}'- i z \blroz_{ss}   & - i z  \blroz_{si}        & i z'  \blroz_{si}'       \\
                - i z \blroz_{is} &  - i z \blroz_{ii} &  \zero{2}  \\
                i z'  \blroz_{is}' & \zero{2}& i z' \blroz_{ii}'
            \end{array}
          \right)
\\
   \vec{M}_1&=&
            \left(
                    \begin{array}{c}
                        -(\dd+\dd') w_p^2- i z \blroz_{s} + i z' \blroz_{s}'  \\
                        -\dd w_p^2 - i z  \blroz_{i}\\
                        -\dd' w_p^2 + i z'  \blroz_{i}'
                    \end{array}
            \right)
\\
\nonumber     \vec{M}_0&=&
        -\frac{w_p^2}{2}\left(\dd^2+\dd'^2\right)- i z \Delta k_z(\kp{s0},\omega_{s};\kp{i0},\omega_{i})\\
        &&+ i z'\Delta k_z(\kp{s0},\omega_{s};\kp{i0}',\omega_{i}')
\end{eqnarray}
We use $\openone_2$ and $\zero{2}$ to denote two dimensional identity and zero matrices, furthermore  $\dd=\V{k}_{s0\perp}+\V{k}_{i0\perp}$, $\dd'=\V{k}_{s0\perp}+\V{k}_{i0\perp}'$ and all primed Taylor series coefficients components $\blroz'_{\mu \nu}$ and $\blroz'_{\mu}$ are taken for the primed idler frequency $\omega_i'$.

\bibliographystyle{apsrev}

\end{document}